\documentclass[10pt, conference, compsocconf]{IEEEtran}
\ifCLASSINFOpdf
\else
\fi
\usepackage{url}
\usepackage{booktabs}
\usepackage{amsmath}
\usepackage{amssymb}
\usepackage{bm}
\usepackage{etoolbox}
\usepackage[utf8]{inputenc}
\usepackage{array,ragged2e}
\usepackage{tikz}
\usepackage{csquotes}
\usepackage{graphicx}
\graphicspath{{./}}
\usepackage[none]{hyphenat}
\usepackage{array}
\usepackage{xspace}
\usepackage{float}
\usepackage{eurosym}
\usepackage[caption=false,font=small,labelfont=bf]{subfig}
\usepackage{color}
\usepackage[justification=centering]{caption}
\usepackage{tkz-euclide}

\usepackage[backend=biber,style= numeric-comp,sorting=none,maxbibnames=3,doi=false,isbn=false,url=false]{biblatex}
\renewbibmacro{in:}{%
	\ifentrytype{article}{}{\printtext{\bibstring{in}\intitlepunct}}}
\usepackage{environ}
\usepackage[linktoc=all]{hyperref}

\usepackage{verbatim}

\usepackage{setspace}

\begin{document}
%
% paper title
% can use linebreaks \\ within to get better formatting as desired
\title{Business Model of a Botnet}
\date{}

\author{\IEEEauthorblockN{C.G.J. Putman}
	\IEEEauthorblockA{University of Twente\\
		Enschede, The Netherlands\\
		Email: c.g.j.putman@student.utwente.nl}
	\and
	\IEEEauthorblockN{Abhishta}
	\IEEEauthorblockA{University of Twente\\
		Enschede, The Netherlands\\
		Email: s.abhishta@utwente.nl }
	\and
	\IEEEauthorblockN{Lambert J.M. Nieuwenhuis}
	\IEEEauthorblockA{University of Twente\\
		Enschede, The Netherlands\\
		Email: l.j.m.nieuwenhuis@utwente.nl}}

% conference papers do not typically use \thanks and this command
% is locked out in conference mode. If really needed, such as for
% the acknowledgment of grants, issue a \IEEEoverridecommandlockouts
% after \documentclass

% for over three affiliations, or if they all won't fit within the width
% of the page, use this alternative format:
% 
%\author{\IEEEauthorblockN{Michael Shell\IEEEauthorrefmark{1},
%Homer Simpson\IEEEauthorrefmark{2},
%James Kirk\IEEEauthorrefmark{3}, 
%Montgomery Scott\IEEEauthorrefmark{3} and
%Eldon Tyrell\IEEEauthorrefmark{4}}
%\IEEEauthorblockA{\IEEEauthorrefmark{1}School of Electrical and Computer Engineering\\
%Georgia Institute of Technology,
%Atlanta, Georgia 30332--0250\\ Email: see http://www.michaelshell.org/contact.html}
%\IEEEauthorblockA{\IEEEauthorrefmark{2}Twentieth Century Fox, Springfield, USA\\
%Email: homer@thesimpsons.com}
%\IEEEauthorblockA{\IEEEauthorrefmark{3}Starfleet Academy, San Francisco, California 96678-2391\\
%Telephone: (800) 555--1212, Fax: (888) 555--1212}
%\IEEEauthorblockA{\IEEEauthorrefmark{4}Tyrell Inc., 123 Replicant Street, Los Angeles, California 90210--4321}}

% use for special paper notices
%\IEEEspecialpapernotice{(Invited Paper)}

% make the title area
	\maketitle
	
	%Main body starts
%Main body starts
\begin{abstract}
	Botnets continue to be an active threat against firms or companies and individuals worldwide. Previous research regarding botnets has unveiled information on how the system and their stakeholders operate, but an insight on the economic structure that supports these stakeholders is lacking. The objective of this research is to analyse the business model and determine the revenue stream of a botnet owner. We also study the botnet life-cycle and determine the costs associated with it on the basis of four case studies. We conclude that building a full scale cyber army from scratch is very expensive where as acquiring a previously developed botnet requires a little cost. We find that initial setup and monthly costs were minimal compared to total revenue.
\end{abstract} 


\section{Introduction and Background}
\label{sec:Introduction}
Botnets and malware over the last couple of years have proven to be a serious threat to cybersecurity. A botnet is a network of various computers which can be controlled by attackers. The controller of the network is called the botmaster. It gives commands to the network by making use of various communication channels. The malicious software used to control this network of computers is known as malware.

It comes as no surprise that the primary motive for the use of botnets is for economic gain \cite{Bottazzi2014}. Although some revenue estimates are known and knowledge regarding the involved actors is available, the structure of the revenue stream is still unclear. In this article we try to provide an insight on the revenue flow, by making use of existing literature and four different case studies.\footnotetext[1]{This paper has been accepted for publication in the proceedings of 2018, 26th Euromicro International conference on Parallel, Distributed, and Network-Based Processing (PDP).}

The practitioners involved in the development botnet can be divided in four tiers \cite{Gosler2013}:

\begin{description}
	\itemsep0em
	\item[Tier 1:] Practitioners who rely on others to develop malicious code, delivery mechanism and execution strategy.
	\item[Tier 2:] Practitioners who have a great depth of experience, with the ability to develop their own tools.
	\item[Tier 3:] Practitioners who focus on the discovery and use of unknown malicious code.
	\item[Tier 4:] Practitioners who are organised, highly technical, proficient and well funded to discover new vulnerabilities and develop exploits.
\end{description}

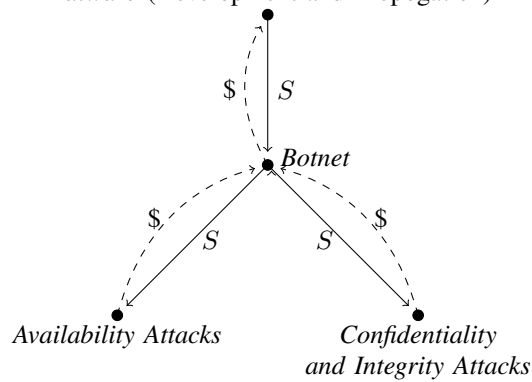
\begin{figure}[h]
	\begin{center}
	\begin{tikzpicture}[scale=1,->, shorten >=5pt]
	\filldraw 
	(0,0)  circle (2pt) node[align=center,above]  {\emph{Malware} (Development and Propogation)}
	(0,-2) circle (2pt) 
	(-2,-4) circle (2pt) node[align=center,   below]  {\emph{Availability Attacks}}
	(2,-4) circle (2pt) node[align=center,   below] (Attacks) {\emph{Confidentiality}\\ \emph{and Integrity Attacks}};
	\draw (0.05,-1.9) node[right]  {\emph{Botnet}};
	\draw (0,0)->(0,-2) node[midway,right] {$S$};
	\draw[dashed] (0,-2) to [bend left] node[midway,left] {\$} (0,0)  ;
	\draw (0,-2)->(-2,-4) node[midway,right] {$S$};
	\draw[dashed] (-2,-4) to [bend left] node[midway,left] {\$} (0,-2)  ;
	\draw (0,-2)->(2,-4) node[midway,left] {$S$};
	\draw[dashed] (2,-4) to [bend right] node[midway,right] {\$} (0,-2)  ;
	\end{tikzpicture}
\end{center}
	$S$: Service Provided 
	\caption{Botnet Ecosystem}
	\label{Fig:Botnet Ecosystem}
\end{figure}

On the basis of this tier distribution we develop a botnet ecosystem as shown in Figure \ref{Fig:Botnet Ecosystem}. In this article we analyse all the linkages of this ecosystem by making use of frameworks like business model canvas, product life cycle analysis and cost benefit analysis.

	\section{Botnet Business model canvas and Life Cycle}
	\label{sec:Building a business model}
	
	\begin{table}[!t]
		\centering
		\resizebox{\linewidth}{!}{
			\begin{tabular}{|c|c| c |c|c|}
				\hline
				\emph{Key Partners} & \emph{Key Activities} & \emph{Value Propositions} & \emph{Customer Relationships} & \emph{Customer Segments}\\
				\hline
				Malware Developers \cite{Bottazzi2014,Miller2010} & Botnet Maintenance \cite{MalwareTech2016,Miller2010} & Advertising products \cite{Kanich2011} & Automation  & Governments \cite{Paganini2013}\\
				Money Handlers \cite{Bottazzi2014,Brunt} & Managing Botnet Infrastructure \cite{Miller2010} & Stealing Money \cite{Bottazzi2014} & Forums \cite{Hilton2016}  & Malicious Parties \\
				Bulletproof Hosting Provider \cite{Bottazzi2014,Brunt} & Perform Attacks \cite{Bottazzi2014,Gosler2013} & Creating Fraudulent ad-clicks \cite{Bottazzi2014} & Internet Relay Chat \cite{Kreing2017} & Website/Server Administrators \cite{Brunt,Krebs2013} \\
				Malware Distributors \cite{Bottazzi2014,Miller2010,Caballero2011} &  & Shutting Down Websites \cite{Brunt,Paganini2013} & Email \cite{Brunt} & \\
				Dark Web Market Places \cite{Radware2016} & & Cryptocurrency Mining \cite{Paganini2013} & & \\
				Government Intelligence Agencies \cite{Paganini2013} & & & & \\
				
				& & & &\\
				\cline{4-4} \cline{2-2} 
				
				& \emph{Key Resources} &  & \emph{Channels} & \\
				
				\cline{4-4} \cline{2-2} 
				& Bots & & Dark Web Marketplaces \cite{Radware2016} & \\
				& Identity Protection Software & & Hacker Forums \cite{Hilton2016} & \\ 
				& Network Connectivity & & Emails \cite{Brunt} & \\
				& & & Websites \cite{Brunt} & \\
				\hline
				\multicolumn{2}{|c|}{\emph{Cost Structure}} & \multicolumn{3}{|c|}{\emph{Revenue Streams}}\\
				\hline
				\multicolumn{2}{|c|}{Malware Development \cite{Bottazzi2014,Miller2010}} & \multicolumn{3}{|c|}{Stolen Bank Account Money \cite{Bottazzi2014}}\\
				\multicolumn{2}{|c|}{Infections \cite{Bottazzi2014,Miller2010,Caballero2011}} & \multicolumn{3}{|c|}{Click Fraud \cite{Bottazzi2014}}\\
				\multicolumn{2}{|c|}{Hosting \cite{Brunt}} & \multicolumn{3}{|c|}{Sale of Booter Services \cite{Brunt,Paganini2013}}\\
				\multicolumn{2}{|c|}{Bandwidth} & \multicolumn{3}{|c|}{Sale of Spam Services \cite{Kanich2011}}\\
				\multicolumn{2}{|c|}{Transaction Fees \cite{Brunt}} & \multicolumn{3}{|c|}{ }\\
				\multicolumn{2}{|c|}{Customer Service \cite{Brunt}} & \multicolumn{3}{|c|}{ }\\
				\hline
			\end{tabular}}
			\captionof{figure}{Business Model Canvas for a Botnet Owner}
			\label{fig:BMC}
		\end{table}
	
	In this paper we use the Osterwalder Business Model Canvas \cite{osterwalder2010business} framework to depict the ``business'' of developing, starting and using a botnet as shown in Figure \ref{fig:BMC}. 
	
	Osterwalder \& Pigneur \cite{osterwalder2010business} propose nine building blocks as the basis of a business model, the logic of how a company intends to generate profit. The nine building blocks, customer segments, value propositions, channels, customer relationships, revenue streams, key resources, key activities, key partnerships and cost structure each have their own core questions that can be used to characterise every business. 
	
	Rodriguez-Gomez et al. \cite{Rodrguez-Gomez2011} proposes are six stages a botnet goes through in its life-cycle: conception, recruitment, interaction, marketing, attack execution and attack success. Understanding the phases of the botnet life-cycle is essential to estimate the costs involved.
	
	\subsubsection{The botnet life-cycle}
	\label{sec:The botnet life-cycle}
	
	The first phase, conception, is all about motivation: why does one want to setup a botnet? On this subject, Rodriguez-Gomez et al. \cite{Rodrguez-Gomez2011} argues that there are five motives for a botmaster to setup a botnet. These are money, entertainment, ego, cause and social status. Of these five it is argued that the primary motive is financial gain. This is usually achieved by selling the source code of the botnet malware. More common is renting out the botnet or its services. Booters are an example of renting services based on botnets \cite{santanna2017thesis}. 
	
	The second stage is the recruitment phase. Infecting computers (or paying others to infect computers for you) with botnet malware resulting in the botmaster being able to control the computer. Usually, larger the botnet the better it is, as the power of a botnet is highly dependent on its size. Depending on the size, renting a botnet for DDoS attacks can cost up to several thousands of dollars a day.
	
	Next, the botmaster can decide to use the botnet himself or rent the services based on botnet. It often takes place by making use of underground online marketplaces or forums, which can be found and accessed via the dark web. In the U.S., the law that prohibits the user to create a botnet (amongst other fraudulent computer activities) is known as the Computer Fraud and Abuse Act \cite{UScode}. Other countries have similar laws in regard to fraudulent computer use, which include botnet use and ownership. 
	
	\begin{table}
		\centering
		\resizebox{0.5\textwidth}{!}{
			\begin{tabular}{c p{2 cm} p{2 cm} p{3 cm} p{3 cm}}
				\hline
				\textbf{Article} & \textbf{Activity} & \textbf{Malware} & \textbf{Context} & \textbf{Finances}\\
				\hline
				\cite{Kanich2011} & Spam advertised pharmaceuticals & Unknown & 360 Million emails per hour 10,000 bots & \$100 per sale, \$3.5 mil. annually\\
				\cite{Bottazzi2014} & Robbing bank credentials & Euro grab-ber, ZeuS based & 30,000 Targets across Europe & \$47 mil. over 2.5 months\\
				\cite{Brunt} & Booter, botnet-for-hire & Unknown & 800,000 Attacks over a 1 year period & \$26k monthly revenue, median of 24 months\\
				\cite{Bottazzi2014} & Advertisement click fraud & Zero-Access & 140,000 hosts & \$900k of daily ad revenue losses\\
				\hline
			\end{tabular}}
			\caption{An overview of botnet case studies}
			\label{table:1}
		\end{table}

	 Bottazzi et al. \cite{Bottazzi2014} states that spamming and DDoS-attacks can be considered least profitable among the activities mentioned in Table \ref{table:1}, since the operation is too noisy, this stands in great contrast to the findings in Kanich et al. \cite{Kanich2011} and the in Brunt et al. discussed vDos case. A summary of previous findings on these cases can be found in Table \ref{table:1}.
	
	\subsubsection{Actors in the life-cycle}
	\label{sec:Actors in the life-cycle}

	Bottazzi et al. \cite{Bottazzi2014} has defined a botnet assembly chain, which has six stages regarding activities varying from development to utilization. Furthermore, this assembly chain is coupled with the level of skill necessary to be able to successfully complete the activity mentioned in the brick, and the “darkness” of the market it is operating in. The latter indicates the level of illegality in the actions performed by the user. Gosler et al. \cite{Gosler2013} propose a similar division, however more based on the involved actors, which he calls tiers. It has already been discussed in Section \ref{sec:Introduction}. 
		
	The first stage of the botnet supply chain, R\&D and Money Transfers are arguably legal businesses. Research and development involves the continuous search for exploits in software, the development of new malicious software and selling knowledge of computer systems and software. The actors behind this process are mostly IT professionals and offer support, customize the software to the wishes of the customer and operate alone or in groups. The actors in this stage of development can be linked to tiers two, three and four in the hierarchy proposed by Gosler et al. \cite{Gosler2013}.
		
	The third and fourth stage, C\&C (command and control) Bulletproof Hosting and PPI (Pay Per Install) distribution operate in a more grey area regarding legality. Bulletproof Hosting includes offering web-based storage for the botnet end-user to store stolen information like banking credentials or passwords on. The hosting service can also serve as the server for the command and control centre of the botnet master. The control centre can consist of one or multiple computers, in the last case typically for redundancy purposes \cite{Rouse2017}. Regarding PPI Distribution; many malware developers do not have the resources to spread the malware they have written to various computers across the world. At least not on a significant scale. To solve this problem they make use of the so called PPI (pay-per-install) Distribution model. In essence this involves the owner of the malware paying affiliates to spread the malware, providing a commission to these malware spreaders per infected device. The client making use of PPI distribution to spread the malware usually collects the funds to be able to afford this by selling regular botnet related services. Taking a look at the tier-based hierarchy, it is likely that the practitioners which are mentioned in tier 1 make use of the PPI Distribution model. Caballero et al. \cite{Caballero2011} indicates the PPI model is one of the most used ways of distributing malware. Estimates are that, of  the twenty most prevalent families of malware, twelve made use of the PPI distribution model \cite{Krebs2011}.
		
		\begin{figure}
			\centering
			\includegraphics[width=0.5\textwidth]{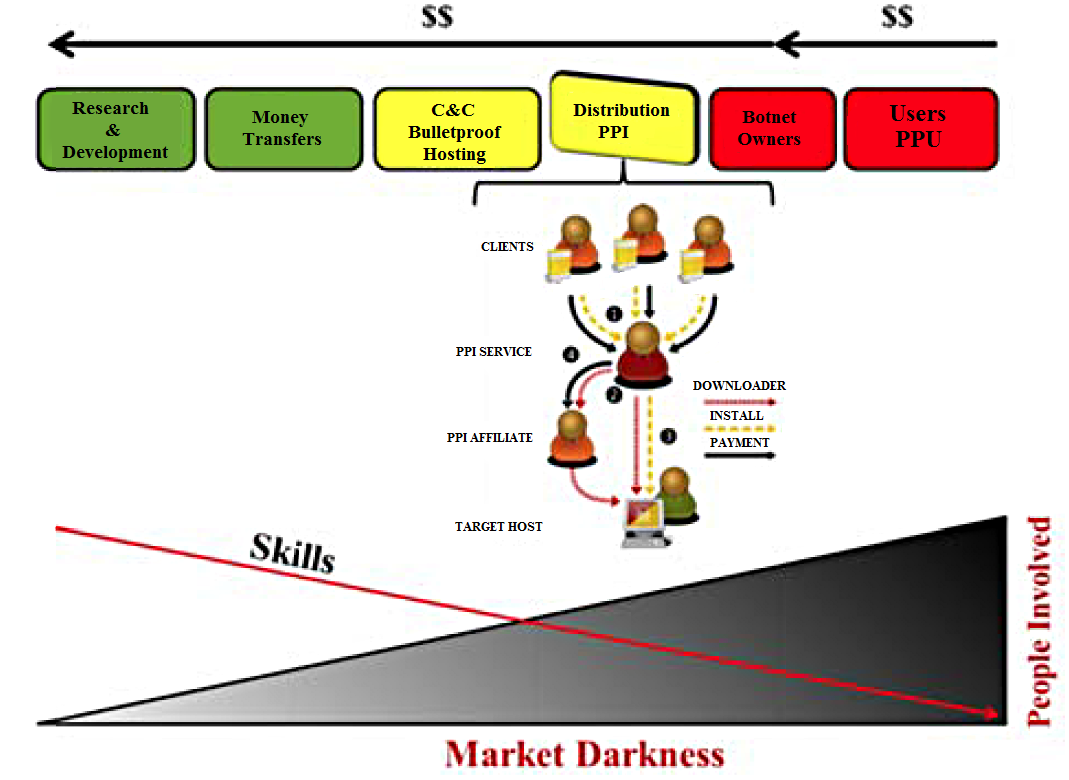}
			\caption{Botnet Assembly Chain \cite{Bottazzi2014}}
			\label{fig:Bottazzi}
		\end{figure}
		
		Lastly, stages five and six enter the indefinite illegal spectrum. Actors involved in this stage are the owners of botnets, the ones who actually perform the attacks.
		
		\section{Life-cycle cost analysis}
		\label{sec: Life-cycle costs analysis }
		By identifying the various phases which occur in the botnet's life-cycle, it should now be possible to estimate the costs involved in each step. In the following sections the life-cycle has been split up in three phases. These steps, acquiring malware, spreading and performing maintenance, form the basis of building the botnet business. 
		
		\subsection{Acquiring malware}
		\label{sec: Acquiring Malware}
		
		\begin{table}
			\resizebox{0.5\textwidth}{!}{
				\begin{tabular}{c c c c c c c}
					\hline
					\textbf{Actors} & \textbf{I.D.I.} & \textbf{Connections} & \textbf{U.S.} & \textbf{NL} & \textbf{U.K.} & \textbf{EU(10 million connections)}\\
					\hline
					Vulnerability analysts & Y & N & 2.9 & 3 & 3.1 & 1.96\\
					Exploit developers & Y & N & 7.3 & 7.6 & 7.8 & 6.55\\
					Bot collectors & N & Y & 4.15 & 0.27 & 1.03 & 0.18\\
					Bot maintainers & Y & Y & 12.9 & 0.88 & 3.42 & 0.48\\
					Operators & Y & Y & 5.4 & 0.37 & 1.4 & 0.2\\
					Remote personnel & Y & N & 0.4 & 0.42 & 0.43 & 0.36\\
					Developers & Y & N & 2.85 & 3 & 3.04 & 2.56\\
					Testers & Y & N & 0.8 & 0.83 & 0.85 & 0.72\\
					Technical Consultants & Y & N & 2 & 2.1 & 2.14 & 1.79\\
					System admins & Y & Y & 0.5 & 0.03 & 0.13 & 0.019\\
					Managers & N/A & N/A & 6.2 & 3 & 3.8 & 2.4\\
					\hline
					TOTAL &  &  & 45.5 & 21.5 & 27.1 & 17.4\\
					\hline
				\end{tabular}
			}
			All costs are in \$ million.
				\caption{Miller's conceptual cyber army \cite{Miller2010}}
				\label{table:2}
		\end{table}
		
		For this calculation\footnote{Miller estimates a total of 540 employees is necessary. One junior manager per ten employees, one senior manager per ten managers. Annual wages are set at \$ 100k respectively \$ 200k.}, a study by Charlie Miller is taken as the baseline \cite{Miller2010}. According to his analysis a full scale act of cybercrime aimed to break down great parts of digital infrastructure of the U.S. will cost the attacker tens of millions of dollars. This therefore indicates such an attack is only really viable for large companies or government like institutions. Furthermore, Miller estimates that planning the attack and developing the malware takes at least a couple of years before it is actually usable. To be able to apply the data he collected into the perspective of other countries we propose the following:
		
		\begin{itemize}
			\item Weigh in the active amount of internet connections of a certain country - (defined as $n$) \cite{SANOU2015}
			\item Weigh in the ITU ICT Development Index (IDI), which is (according to ITU) “used to monitor and compare developments in information and communication technology (ICT) between countries and over time” (defined as $i$) \cite{ITU2015}
			\item Set the active amount of internet connections and IDI base level at the level of the United States.
		\end{itemize}
		
		This has resulted in the following equation:
		
		\begin{equation}
		\label{miller1}
		\dfrac{n}{n_{us}}\times\dfrac{i}{i_{us}}\times Miller's \ estimated \ costs
		\end{equation}
		
		In order to apply this formula correctly, it is important that both the number of active internet connections and the IDI are estimated in the same year. The costs estimated by Miller have been released in 2010. Unfortunately, we have not been able to find conclusive data regarding the IDI and the number of connections from this year. Therefore, In this study, we make use of data collected in 2015. 
		
		The above calculations in case of the Netherlands result in an estimate of annual costs of \$3.11 million to be able to launch a full scale attack on the internet infrastructure of the Netherlands. This calculation is, however, a very rough and an unreliable estimate. Fixed and  managerial costs should be different (in case of Netherlands) than the ones defined by Miller. For example, spreading costs will most likely be a lot less when compared to the U.S., simply because the amount of devices is significantly lower. Therefore, we propose to apply the calculation above only on certain aspects of malware development. We show these calculations in the case of three different countries in Table \ref{table:2}. 
		
		Harris et al. \cite{Harris2013} argues that for “the true do it yourself type” there are botnet malware \emph{how to} guides  available for less than \$600. This seems logical: the mentioned ZeuS and Mirai botnet packages are already set up, but still need some configuration to get them running. The more expensive ZeuS packages offer easier and more complete functionality.
		
		\subsection{Spreading the malware}
		\label{sec: Spreading the malware}
		 In Miller’s case, if one would want to calculate the total costs of malware, including spreading but excluding the use and maintenance of the botnet, it should be possible to add up \$160k to the \$16 million which is already accounted for. In other cases, the average botmaster will most likely make use of the earlier describe paid per install model. According to Brian Krebs, making use of the PPI distribution model, costs are estimated to vary from \$7 to \$180 per 1000 installations \cite{Krebs2011}. As both researches have been conducted around the year 2011, it should be possible to compare these costs. Taking the average of \$93.5 per 1000 installations (\$0.0935 per installation), and comparing this to the estimated \$160k spreading costs for an average EU country, with the PPI distribution model it would be possible to infect over 1.7 million devices. The spreading costs in Miller’s report appear to be more economical, at around \$0,016 per infected device.
		
		\subsection{Botnet maintenance}
		\label{sec: Botnet Maintenance}
		Botnet malware has to be maintained to ensure effectiveness. In the case the owner of the botnet is also the author of the malware code, he can resolve these issues himself and release new revisions of his malware if necessary. According to a Mirai based case study, maintenance costs of desktop botnets exceeds the revenue generated by DDoS attacks \cite{MalwareTech2016}. With the rise of IoT, a new target group consisting of less secure devices like CCTV cameras has risen. A botnet composed of these less secure devices may be a lot cheaper to maintain even though it is ten times bigger. The Mirai botnet makes use of these devices. Although the malware can be cleared off an infected device by simply rebooting it (as the malware is only stored on the device’s RAM, which gets erased when rebooting) the same device can most likely be re-infected almost immediately \cite{Hamilton2016}. The botmaster only has to keep a list of IP-addresses which have been infected in the past, and scan if these IP-addresses are still connected to the botnet.
		
		Re-infection costs have been estimated at \$0.0935 per device. Furthermore, the malware creator has to constantly innovate, as security updates are rolled out every day. Lastly, in some cases the malware creator provides customer service to his customers, which bought a malware package from him. According to Miller, a level 1 malware developer earns around \$125k a year. Assuming the specific malware is being developed and maintained by a level 1 developer, maintaining the botnet costs the developer around \$59 per hour. (Based on 22 working days a month, working 8 hours a day). Let’s apply these costs to the two botnet malware types:
		
		\textbf{ZeuS} - As re-infection is impossible, the botmaster has to rely on unpatched devices. If this is the case each infection costs the estimated \$0.0935. According to ZeuS tracker, the average ZeuS binary antivirus detection rate lies around 40\%. Therefore, finding unpatched devices should not be a problem. If the problem should arise new malware has to be developed, or the ZeuS malware has to be altered, this would cost the developer the estimated \$59 per hour. Assuming the developer has a normal tax-paying 8 hour a day job besides malware developing, 8 hours of sleep a day and some spare time, the developer can spend around 4 hours a day on developing and/or maintenance. Over a year’s period, instead of maintaining the botnet, the developer could have worked a job which would have provided him around \$62k. This will be regarded as maintenance costs per year. 
		
		\textbf{Mirai} - As the owner of the infected device has very little influence in making sure his/her device gets patched, or requires at least some advanced computing knowledge to do so, one can assume successful re-infection is more likely to occur than not. Assuming the 2 year warranty period as the minimum lifespan of the product, chances of the device being replaced by a newer non vulnerable device (assuming the device owner does not know his device has been infected) seems negligible. This makes maintaining a Mirai botnet much easier and cost efficient, as each re-infection (if done by making use of the PPI model) only costs the estimated \$0.0935. This is, however, all speculation as conclusive data regarding the maintenance of Internet of Things botnets is currently unavailable.
		
		Depending on the intended use of the botnet, it can now be used or marketed to rent it out. In both cases the botnet is being run from one or multiple command and control centres, which are basically computers which run the software used to control the botnet. Controlling the botnet costs time, of which the costs per hour have been set at \$59 in the previous section. Marketing and selling the botnet costs little to no money.
		
		\begin{table}
			\centering
			
			\resizebox{0.5\textwidth}{!}{
				\begin{tabular}{c c c c}
					\hline
					& \textbf{ZeuS} & \textbf{Mirai} & \textbf{Miller(10 million EU connections)}\\
					\hline
					Malware package & \$700 up to $<$\$10k & $<$\$30 & N/A\\
					Malware development & \$125k & Unknown & \$16 million\\
					Spreading per device & \$0.0935 & \$0.0935 & \$0.016\\
					Maintenance & \$62k & Unknown & \$48k\\
					Marketing & \$28.8k & \$28.8k & N/A\\
					\hline
				\end{tabular}
			}
			\caption{Estimated botnet setup costs.}
			\label{table:3}
		\end{table}
		
		Summarizing, the total setup costs for a ZeuS or Mirai based botnet can be found in Table \ref{table:3}. All numbers are on annual basis, except of malware package and spreading costs.
		
		\section{Botnet economics}
		\label{sec: Botnet economics}
		The information given in Section \ref{sec: Life-cycle costs analysis } provides an adequate basis to estimate money that flows to the various actors that occur in the life-cycle.
		
		\subsection{Cost-benefit analysis}
		\label{sec: Cost-benefit analysis}
		
		\begin{table}
			
			\centering
			\resizebox{\linewidth}{!}{
				\begin{tabular}{p{2 cm}| c c c c |p{3 cm}}
					\hline
					& DDoS - 30000 bots \footnote{vDos claimed attacks with power up to 216Gbps. Average internet 
						speed around the world lies at 7.2Mbps. \cite{globalinternetspeed}\cite{VDosStresser2014}} & Bank Fraud - 30000 bots & Spamming - 10000 bots & Click fraud - 140000 bots & \\
					\hline
					N/A & Unknown & ZeuS & Unknown & ZeroAccess & Malware type\\
					Developer & Mirai: \$30 Zeus: \$700 & \$700 up to $<$ \$10k\footnote{Monthly recurring costs, price varies between packages. \cite{Neville2013}} & Unknown & \$5k up to 10k & Malware package costs\\ 
					Money handler & \$780 & \$564000 & \$9000 & \$750000 & Transaction fees at 3\% \\
					(Bulletproof) web hosting provider & \$2400 & $<$ \$70 & $<$ \$2400 & $<$\$70 & Web and C\&C bulletproof hosting costs \\
					Distributor & \$2805 & \$2805 & \$935 & \$13090 & Distribution costs by the PPI model \\
					Botmaster/user & \$26000 & \$18.8 million & \$300000 & $<$\$25 million & Monthly revenue \\
					\hline
					\emph{Monthly Profit} & ZeuS: \$19315 Mirai: \$19985 & $>$\$18.2 million & $>$\$287665 & $<$\$24.2 million & \\ 
					\hline
				\end{tabular}
			}
			\caption{Aggregated botnet cost/benefits based on case studies}
			\label{table:4}
		\end{table}
		While the analysis in Section \ref{sec: Botnet Maintenance} focuses on ZeuS and Mirai, the analysis that follows is activity based. The four main botnet related activities, DDoS attacks, spamming, bank fraud and click fraud are linked to a specific type of malware. For example, as explained before, Mirai is only suitable for DDoS attacks \cite{Akamai2007}. A detailed overview of the researched cases can be found in Table \ref{table:4}.  
		
		Since initial costs vary little between each case compared to the potential profits, it is no wonder spreaders and hosting providers take a significant part of DDoS revenue. Around 25\% of total revenue per month to be precise, assuming every bot has to be re-infected every month. A significant part, especially when compared to the maximum of 1.1\% of hosting and spreading costs of the other three cases. Hosting could be seen as a viable business, as it brings relatively low risk with it. Spreading, however, brings great risks as it is demonstrable punishable by law. The costs that have to be made to setup a botnet are in most cases insignificant. It seems, however, the real winners are the money handlers.
		
		\subsection{Economic impact on organisations}
		\label{sec: Economic impact on institutions}
		
		Information regarding the overall economic impact of botnet attacks on various institutions is scarce. A possible cause for this could be that it is hard to estimate the economic impact: it is not always known which institutions are affected, and economic impact is more often than not an indirect impact \cite{jowua17-8-4-01,AbhishtaJN17}. 
		
		A 2011 report of intelligence provider Detica on the impact of cybercrime on U.K. based institutions defines four types of costs associated with cybercrime: costs in anticipation of, costs as a consequence of, costs in response to and indirect costs associated with cybercrime \cite{Report2011}. Regarding direct and indirect impact, Anderson et al. elaborates on the aspects mentioned above, and give a few examples, based on bank fraud \cite{Anderson2013}. Table \ref{Table:5} describes for each type of attack the subsequent costs for the target company.
		
		\begin{table}
			\centering
		
			\resizebox{0.5\textwidth}{!}{
				\begin{tabular}{c c c c}
					\hline
					\textbf{Cybercrime type} & \textbf{Percentage of costs} & \textbf{Annual costs (in million \$)} & \textbf{Attack costs (in \$ $\times$1000)}\\
					\hline
					Malware & 16.50\% & 1.57 & 5.11\\
					Phishing / SE & 13\% & 1.24 & 95.8\\
					Web-based attacks & 16.70\% & 1.59 & 88.1\\
					Malicious code & 12.50\% & 1.19 & 92.3\\
					Botnets & 2.80\% & 0.27 & 0.995\\
					Stolen devices & 8.80\% & 0.84 & 31.9\\
					DDoS & 17.50\% & 1.66 & 133.5\\
					Malicious insiders & 12.20\% & 1.16 & 167.9\\
					\bottomrule
				\end{tabular}}
					\caption{Financial impact of botnets on institutions}
					\label{Table:5}
			\end{table}

			\section{Conclusions}
			\label{sec: Conclusions}
			In this paper we present the botnet business model with the help of the canvas proposed by \citeauthor{osterwalder2010business}\cite{osterwalder2010business}. We then discuss the stages and actors involved in a botnet life cycle in Section \ref{sec:Building a business model} and analyse the life cycle costs of a botnet in Section \ref{sec: Life-cycle costs analysis }. Finally, in Section \ref{sec: Botnet economics} we perform a cost benefit analysis from the view point of an attacker on the basis of four case studies and discuss the financial impact of botnets on organisations. We draw the following conclusions from our analysis:  
		 
			\begin{itemize}
				\item By making use of readily available malware packages and outsourcing malware infection, it is possible to set up a profitable botnet business within days.
				\item Initial investment costs on acquiring malware, spreading, combined with recurring costs like hosting and transaction fees, are nearly insignificant.
				\begin{itemize}
					\item In three out of four researched cases, set-up costs accounted for a maximum of 1.1\% of monthly revenue.
				\end{itemize}
				\item Profitability between various botnet uses varies drastically. DDoS-for-hire (Booter) is the least profitable, but also has the longest life time.
			\end{itemize}	
			With the help of gathered knowledge regarding botnet actors and revenue streams, future research should focus on intercepting the revenue streams of botnets.
			
			\printbibliography
		
		\end{document}